\documentclass[11pt,a4paper]{article}
\usepackage{amsfonts}
\usepackage{graphicx}
\usepackage{amsmath}
\usepackage{hyperref}
\usepackage{enumerate}
\usepackage{amsmath,amssymb}
\usepackage{slashed,mathrsfs}
\usepackage{feynmp,subfigure}
\usepackage{verbatim,graphicx}
\usepackage[sub,ovp]{psfragx}
\usepackage{overpic}

\RequirePackage{mathrsfs} \RequirePackage[sc]{mathpazo}
\RequirePackage{wasysym} \RequirePackage{setspace}
\newcommand{\be}{\begin{equation}}
\newcommand{\ee}{\end{equation}}
\newcommand{\bea}{\begin{eqnarray}}
\newcommand{\eea}{\end{eqnarray}}
\input{tcilatex}
\begin{document}

\date{ }
\title{ \rightline{\mbox{\small
}}\textbf{\ On Gravity Implication in the Wavefunction Collapse }}
\author{A. Belhaj$^{1}$, S. E. Ennadifi$^{2}$\thanks{%
Authors in alphabetical order} \thanks{%
Corresponding author: ennadifis@gmail.com} \\
%EndAName
\\
{\small $^{1}$Physics department, Faculty of Sciences, Mohammed V University
in Rabat, Rabat, Morocco }\\
{\small $^{2}$LHEP-MS, Faculty of Sciences, Mohammed V University in Rabat,
Rabat, Morocco}\\
}
\maketitle

\begin{abstract}
Inspired by an ontic view of the wavefunction in quantum mechanics and
motivated by the universal interaction of gravity, we discuss a possible
gravity implication in the state collapse mechanism. Concretely, we
investigate the stability of the spatial superposition of a massive quantum
state under the gravity effect. In this context, we argue that the stability
of the spatially superposed state $\left\vert \Psi \right\rangle =$ $%
\sum\limits_{i=1}^{n}\alpha _{i}\left\vert \psi _{i}\right\rangle $, $\alpha
_{i}\in \mathbb{C}$, depends on its gravitational self-energy $U_{\Psi }^{G}$
originating from the effective mass density distribution $\rho _{m}^{\ast
}\left( x\right) $ through the spatially localized eigenstates $\left\vert
\psi _{i}\right\rangle $. We reveal that the gravitational self-interaction
between the different spacetime curvatures $\left\vert G_{i}\right\rangle $
created by the eigenstate effective masses $m_{\psi }^{\ast }$ leads to the
reduction of the superposed state to one of the possible localized states $%
\left\vert \psi _{k}\right\rangle \left\vert G_{k}\right\rangle $. Among
others, we discuss such a gravity-driven state reduction. Then, we approach
the corresponding collapse time $\tau _{\Psi }$ and the induced effective
electric current $I_{\Psi }^{\ast }$ in the case of a charged state, as well
as the possible detection aspects.


%\textit{PACS}:\ bear this framework in mind here,
\end{abstract}

\newpage

\section{Introduction}

It is now mostly accepted that the physical properties of nature at the
level of atoms and subatomic particles are well-characterized by Quantum
Mechanics (QM), which is a succeeded basic physical theory \cite{1,2,3}. It
is the ground of all quantum branches of physics, from fundamental quantum
sciences to quantum technologies. Yet, since the outset of QM oneself the
so-called problem of measurement still preoccupies the minds of physicists
and philosophers from decenniums. The problem is to understand how the
observed classical world releases from the fundamental QMcal one. In spite
of the progressions made, remarkably across the idea of the decoherence, the
problem of measurement persists. Seeing that such a fundamental issue stays
unsettled for decades of efforts is an indication for the incompleteness of
QM as a physical theory \cite{4,5,6}. In QM, the problem of measurement is
associated with the willingness of establishing meaning of the state vector
reduction scenario, i.e., the wavefunction collapse. This problem takes
place due to the fact that the Schr\"{o}dinger equation alone is incapable
to predict the lucidity of a singular outcome in a quantum measurement
experiment. Rather, it predicts the existence of the superposed
macroscopically different states encompassing all possible outcomes
simultaneously \cite{7,8,9}. Diverse interpretations have been suggested to
evince the obscurity of the measurement. These are essentially Copenhagen
interpretation, projection postulate, many-worlds, and ontic quantum state 
\cite{8,10,11,12,13}. Another possible road, therewith, is to prevent the
occurence of superpositions by adjusting the dynamics of QM. In particular,
this could be done by complementing the Schr\"{o}dinger equation with minor
non-linear and stochastic terms, such as the Ghirardi--Rimini--Weber theory
(GRW) and the Continuous Spontaneous Localisation (CSL) collapse theories 
\cite{14,15,16,17,18}. In the same path, the spontaneous collapse event has
been thought to be linked to gravity, as it is the case in the Di\'{o}%
si--Penrose model where it is suggested that the integration of gravity
could sort out such foundational issues and offer an entire physical theory 
\cite{19,20}.

In this work, guided by an ontic view of the QMcal wavefunction, we
contribute to these endeavors by discussing a possible gravity implication
in the reduction process of the quantum state. For that, we deal with the
stability behavior of the spatial superposition of a quantum state in QM
with the presence of the gravitational interaction. After a concise
presentation of the state of a quantum system and the related measurement
problem, we first consider the spatially superposed state $\left\vert \Psi
\right\rangle =$ $\sum\limits_{i=1}^{n}\alpha _{i}\left\vert \psi
_{i}\right\rangle $, $\alpha _{i}\in \mathbb{C}$, under its gravitational
self-energy $U_{\Psi }^{G}$ originating from the effective mass density
distribution $\rho _{m}^{\ast }\left( x\right) $ through the spatially
localized states $\left\vert \psi _{i}\right\rangle.$ Precisely, we
investigate the state reduction driven by the gravitational self-interaction
between the different spacetime curvature states $\left\vert
G_{i}\right\rangle $ created by the localized state effective masses $%
m_{\psi _{i}}^{\ast }$ leading to the decay of the quantum state $\left\vert
\Psi \right\rangle $ to one of the possible localized states $\left\vert
\psi _{k}\right\rangle \left\vert G_{k}\right\rangle $ after a certain
interaction time. In terms of the gravitational self-energy $U_{\Psi }^{G}$,
we then approach the corresponding collapse time $\tau _{\Psi }$ and the
induced effective electric current $I_{\Psi }^{\ast }$ for charged quantum
states. We also discuss the possible detection aspects. We end with a
conclusion and provide further considerations.

\section{Quantum state and measurement problem}

In the current formulation of quantum theory, every state of a physical
system, denoted by $\left\vert \Psi \right\rangle $, is representable by a
vector in an abstract complex space called Hilbert vector space. Vectors
belonging to a given family differ only in phase contributions. The dynamics
of the quantum system is described by the evolution of the state vector
according to the following Schr\"{o}dinger equation 
\begin{equation}
i\hbar \partial _{t}\left\vert \Psi \right\rangle =\widehat{H}\left\vert
\Psi \right\rangle  \label{eq1}
\end{equation}%
where $\widehat{H}$ is the Hamiltonian of the system. Such an evolutionary
equation is linear, deterministic, and unitary. The dynamical quantities,
represented by linear Hermitian operators acting on the state vector, are
called observables $\widehat{O}$ and whose eigenvectors $\left\vert \psi
_{i}\right\rangle $ being particular solutions to the Schr\"{o}dinger
equation (\ref{eq1}) with certain measuring chance probabilities. The
eigenvectors $\left\vert \psi _{i}\right\rangle $ form a complete
orthonormal basis for the corresponding Hilbert space. In terms of such an
arbitrary system, the state $\left\vert \Psi \right\rangle $ can always be
expressed as a linear combination as follows 
\begin{equation}
\left\vert \Psi \right\rangle =\sum\limits_{i=1}^{n}\alpha _{i}\left\vert
\psi _{i}\right\rangle  \label{eq2}
\end{equation}%
where $\alpha _{i}$ are complex numbers satisfying the normalization
condition $\sum\limits_{i=1}^{n}\left\vert \alpha _{i}\right\vert ^{2}\text{%
\ }=1$, and where $n$ is the dimension of the corresponding Hilbert space.
According to this, every combination of all the eigenstates of the Schr\"{o}%
dinger equation (\ref{eq1}) governing the system is also a state of the
system. This is referred to as the superposition principle. This principle
immediately drives to the problem of measurement, i.e., what will be the
result of a measurement on such superposed states (\ref{eq2})? If the
measurement action obeys the standard pattern, the result can only be one of
the possible eigenstates and it can not be any mixture of them. The answer
is upon the measurement of an observable $\widehat{O}$ on a superposed state 
$\left\vert \Psi \right\rangle $. The result will be randomly one of the
eigenstates $\left\vert \psi _{k}\right\rangle $ with eigenvalue $o_{k}$
with a chance, i.e., the probability\ $P\left( o_{k}\right) =\left\vert
\alpha _{k}\right\vert ^{2}$. According to the Born rule \cite{1}, the state
of the system, immediately, after the measurement will be $\left\vert \Psi
\right\rangle ^{\prime }$ such as 
\begin{equation}
\left\vert \Psi \right\rangle ^{\prime }=\text{\ }\left\vert \psi
_{k}\right\rangle =\frac{\widehat{P}_{k}\left\vert \Psi \right\rangle }{%
\sqrt{P\left( o_{k}\right) }}.  \label{eq3}
\end{equation}%
Here, $\widehat{P}_{k}=$ $\sum\limits_{\ell =1}^{g_{\ell }}\left\vert
u_{k,\ell }\right\rangle \left\langle u_{k,\ell }\right\vert $\ is the
projection operator on the propre subspace associated with the state $%
\left\vert \psi _{k}\right\rangle $ where $g_{\ell }$ is the degree of
degeneracy of the eigenvalue $o_{k}$. This is known as the \emph{wave packet
reduction} postulate. According to the dominant QMcal view, the measurement
process takes place during the interaction time $t=\tau $ of the measuring
apparatus $A$ with the quantum system in a state $\left\vert \Psi
\right\rangle $. Concretely, due to the linear nature of QM, this can be
summarized by the following evolution apparatus-system state equation 
\begin{equation}
\left\vert \Psi \right\rangle \left\vert A_{0}\right\rangle =\left(
\sum\limits_{i=1}^{n}\alpha _{i}\left\vert \psi _{i}\right\rangle \right)
\left\vert A_{0}\right\rangle \overset{\tau }{\rightarrow }\left(
\sum\limits_{i=1}^{n}\alpha _{i}\left\vert \psi _{i}\right\rangle \right)
\left\vert A_{i}\right\rangle \text{\ }  \label{eq4}
\end{equation}%
where $\left\vert A_{0}\right\rangle $ represents the initial state of the
apparatus and $\left\vert A_{i}\right\rangle $ denotes its orthogonal final
states (assumed to correspond to the observer perceptions). This whole
evolution equation (\ref{eq4}) displays clearly an entanglement between the
quantum and the apparatus systems. Thus, the latter can not be claimed to
own a determined macroscopic state corresponding to a possible definite
observation. According to this, the fundamental problem is to give sens to
the reduction of the superposed state vector (\ref{eq4}) towards the
observed classical one. Certain procedures have been proposed in order to
explain such a breakdown of the quantum state superposition. In addition to
the most influential \emph{Copenhagen interpretation} \cite{1,10,12}, there
are the \emph{incompleteness}, the \emph{decoherence} or \emph{%
environment-entanglement} \cite{21}, and the \emph{many world interpretation}
\cite{22}. Alternative interpretations, where a different reformulation of
the standard quantum procedures, have been conducted \cite%
{23,24,25,26,27,28,29}. According to such suggestions, there are ones
putting that deviations from standard unitary evolution are to be found in
the behavior of conscious observers. In a more physical vision, however, the
proposed deviations, from precise unitary evolution, become perceptible. In
some appropriate sens, they behave simply when the scale of the system
becomes large. This scale largeness does not necessary mention only the
number of particles of the physical system \cite{14}, but it might also
mention its mass distribution \cite{19,20}. In the latter scheme, it is
naturally considered that it is gravity being behind the effect that allows
deviations from ordinary quantum laws. In this view, and for more
generality, it is prospective to let the scale largeness of the system
refers to its energy distribution and investigate the possible influence of
gravity.

\section{Spatial superposition and gravitational state reduction}

\subsection{Emergence of gravitational field states}

To start, we assume that a spatial quantum superposition of many states has
been established, where each single state $\left\vert \psi _{i}\right\rangle 
$ has a well-determined energy, where the energies differ from one state to
the other. Owing to the effects of these distributed different energies on
the spacetime according to General Relativity (GR), the physical geometries
associated with all the several single state locations are now all distinct
from one to another. In accordance with this, each single state position is
joined with the gravitational field state $\left\vert G_{i}\right\rangle $.
These states refer to the QMcal description of the macroscopic gravitational
fields, i.e., gravitational field quantum states $\left\vert
G_{i}\right\rangle $, created by the mass energies of the single states $%
\left\vert \psi _{i}\right\rangle $ at their locations. Thus, the spatially
superposed state must be envisaged now as an entangled one as it will be
shown in what follows. In fact, in a different reasoning concerning the
introduction of the underlying gravitational field from the one given in the
Di\'{o}si--Penrose model \cite{19,20,30} where the description of the
involved gravitational interaction is introduced in terms of a classical
noise field related to the Newtonian potential or the associated induced
spacetime curvatures, we propose here a possible way to bring out the
corresponding gravitational states from the initial superposed state via the
complex coefficients $\alpha _{i}$\ that should be treated as functions of
the spacetime position $x$. More precisely, we could express them in terms
of their real and angular local phase (spacetime dependent) parts in the
following manner 
\begin{equation}
\alpha _{i}=\alpha _{i}\left( x\right) =\alpha _{i}\left[ \theta \left(
x\right) \right] =\left\vert \alpha _{i}\right\vert e^{i\theta _{i}\left(
x\right) }:=\alpha _{i}^{G}e^{i\theta _{i}\left( x\right) }  \label{eq5}
\end{equation}%
where $\alpha _{i}^{G}$ are the real parts and $\theta _{i}\left( x\right) $
are the local phases of the angular part. By considering these local angular
part functions as wavefunctions 
\begin{equation}
e^{i\theta _{i}\left( x\right) }=e^{i\theta _{i}\left( x\right)
}:=G_{i}\left( x\right) ,  \label{eq6}
\end{equation}%
the spatially superposed state takes now the expanded form 
\begin{equation}
\left\vert \Psi \right\rangle =\sum\limits_{i=1}^{n}\alpha
_{i}^{G}\left\vert \psi _{i}\right\rangle \left\vert G_{i}\right\rangle ,%
\text{\ }\sum\limits_{i=1}^{n}\left( \alpha _{i}^{G}\text{\ }\right) ^{2}=1.
\label{eq7}
\end{equation}%
In this way, the gravitational field states $\left\vert G_{i}\right\rangle $
could be associated with the local phases of the angular parts of the
complex coefficients. In this regard, it is recalled that once we permit
such a spacetime-dependent phase transformation through (\ref{eq6}), the
corresponding Schr\"{o}dinger equation is no longer invariant, since one has
the deviated transformation law

\begin{equation}
\partial _{\mu }\Psi =\partial _{\mu }\left( \psi G_{\psi }\right) =\partial
_{\mu }\left( \psi e^{i\theta }\right) =e^{i\theta }\left( \partial _{\mu
}\psi +i\psi \partial _{\mu }\theta \right) \text{\ }  \label{eq8}
\end{equation}%
where the first term represents the standard coordinate changing of the
wavefunction, while the second one represents a correction describing the
evolution or the deformation of the coordinate system with respect to such a
derivative. In GR, the manner to render the transformation with local phase (%
\ref{eq8}) a symmetry then again is to introduce a spacetime geometric
structure term (as corrections to an ordinary derivative in flat spacetime),
i.e., affine connection $\Gamma _{\mu \nu }^{\rho }$ which transforms under
a change of coordinates according to the rule $\Gamma _{\mu \nu }^{\rho
}\equiv \frac{\partial x^{\rho }\partial x^{k}\partial x^{\lambda }}{%
\partial x^{\sigma }\partial x^{\mu }\partial x^{\nu }}\Gamma _{k\lambda
}^{\sigma }+\frac{\partial x^{\rho }}{\partial x^{\sigma }}\frac{\partial
^{2}x^{\sigma }}{\partial x^{\mu }\partial x^{\nu }}$. Once this is done,
the\ above ordinary partial derivative transformation can be replaced by the
covariant one

\begin{equation}
\triangledown _{\mu }\Psi \equiv \left( \partial _{\mu }-\delta _{\rho
}^{\nu }\Gamma _{\mu \nu }^{\rho }\right) \Psi .  \label{eq9}
\end{equation}%
By considering the phase transformation $e^{i\theta }$ and the second term
in (\ref{eq8}) analogous to the basis transformation $\frac{\partial x^{\rho
}}{\partial x^{\sigma }}$ and the extra $\frac{\partial ^{2}x^{\sigma }}{%
\partial x^{\mu }\partial x^{\nu }}$ in $\Gamma _{\mu \nu }^{\rho }$,
respectively, the \emph{unwanted} terms in the transformation of e.g. $%
\partial _{\mu }\Psi $ and $\Gamma _{\mu \nu }^{\rho }$ cancel by performing
a coordinate transformation. Thus, we have the nice transformation $%
\triangledown _{\mu }\Psi \rightarrow e^{i\theta }\Psi $ where the required
invariance is restored.

\subsection{Gravitational self-energy and state reduction mechansim}

Now, the key idea is that these generated gravitational field states are 
\emph{disliked} by Nature and thus will be suppressed since the spacetime
metric should be well defined. Such a suppression process corresponds to the
reduction of the superposed state into one of the localized states.
Actually, according to the non-linear effect of GR, the gravitational
interaction effect between any pair of these superposed spacetime location
states (\ref{eq6}) would have to be taken into account. Therefore, there
will be a considerable attraction of parts of the closer spatial
superpositions. This gives rise to a generation of position and momentum
correlations leading to the increase of quantum entanglement. Consequently,
these spacetime states $\left\vert \psi _{i}\right\rangle \left\vert
G_{i}\right\rangle $ would \emph{fall} towards each other in accordance with
their gravitational attractions, corresponding to a non-static spacetime
geometry. In this regards, we should now consider carefully what this means
in a context such as this. Here, without giving a precise dynamical equation
for the collapse as was done by Di\'{o}si \cite{19}, we follow the Penrose's
line of reasoning \cite{20,30} but with a different conservative approach
consisting of a description of the collapse mechanism in terms of spatial
translations scenario and use the mass density concept to determine the
underlying gravitational self-energy, and then derive the corresponding
decay rate. Concretely, since there is a spatial translation $%
x_{i}\rightarrow x_{i}-a_{k_{i}}\equiv x_{k}$\ of one spacetime state
location $\left\vert \psi _{i}\right\rangle \left\vert G_{i}\right\rangle $\
to another $\left\vert \psi _{k}\right\rangle \left\vert G_{k}\right\rangle $%
\ under the effect of the spacetime curvatures generated by the effective
mass energies $m_{\psi _{i}}$, let $\widehat{T}\left( a_{k}\right) _{k=1..n}$%
\ be the operator acting on the $n$\ localized states translating them into
one state. In this scenario, and for infinitesimal spatial translations,
such a translational operator could be expressed as 
\begin{equation}
\widehat{T}\left( a_{k}\right) _{k=1,\dots ,n}=e^{-a_{k}^{\mu }\nabla _{\mu
}}  \label{eq10}
\end{equation}%
which when it acts on a state centered at position $x_{i}$ moves it to the
one centered at position $x_{k}$. Concretely, we have 
\begin{equation}
\widehat{T}\left( a_{k}\right) \left( \left\vert \psi _{i}\right\rangle
\left\vert G_{i}\right\rangle \right) =\left\vert \psi \left(
x_{i}-a_{k_{i}}\right) \right\rangle \left\vert G\left(
x_{i}-a_{k_{i}}\right) \right\rangle \equiv \left\vert \psi
_{k}\right\rangle \left\vert G_{k}\right\rangle .\text{\ \ }\   \label{eq11}
\end{equation}%
Although for QM one generally needs to study full quantum field theory in a
classical curved spacetime, the corresponding QMcal momentum operator
generating the gravitationally curved path for such a spacetime geometry
state is $\widehat{p}_{\mu }=-i\hbar \nabla _{\mu }=-i\hbar \left( \partial
_{\mu }+\frac{1}{2}\frac{\partial _{\mu }\sqrt{g}}{\sqrt{g}}\right) $, where 
$g=\left\vert \det g\right\vert $ is the metric of the spacetime geometry
under consideration. Based on such arguments, the translational operator (%
\ref{eq10}) reads now as 
\begin{equation}
\widehat{T}\left( a_{k}\right) _{k=1,\ldots ,n}=e^{-\frac{i}{\hbar }%
a_{k}^{\mu }\widehat{p}_{\mu }}.  \label{eq12}
\end{equation}%
With this description beard in mind, and as it has been already mentioned
above, all the spacetime geometry states will be subject to the spacetime
translation (\ref{eq11}) under the involved gravitational self-attractions
emerged from the spacetime curvature effects. As a result of this
gravitational self-interaction, the initial superposed state (\ref{eq7})
will undergo a $n-1$ translational state reduction process into one of the
single localized states after a certain interaction time $\tau _{\Psi }$
such as

\begin{equation}
\prod\limits_{r=1}^{n-1}\widehat{T}_{r}\left\vert \Psi \right\rangle
=\prod\limits_{r=1}^{n-1}\widehat{T}_{r}\left[ \sum\limits_{i=1}^{n}\alpha
_{i}^{G}\left\vert \psi _{i}\right\rangle \left\vert G_{i}\right\rangle %
\right] \equiv \left\vert \psi _{k}\right\rangle \left\vert
G_{k}\right\rangle .  \label{eq13}
\end{equation}%
The resulting single state corresponds to the observed one after the
gravitational self-interaction process with a probability amplitude $\alpha
_{k}^{G}$. The latter corresponds to the probability amplitude of finding
the initial unstable superposed state system $\left\vert \Psi \right\rangle $
in the single state $\left\vert \psi _{k}\right\rangle \left\vert
G_{k}\right\rangle $ after the interaction time $\tau _{\Psi }$. From
related theoretical results \cite{16,19,20,30,31,32}, the gravitational
reduction process is determined by the gravitational self-energy of the
system in this quantum state arising from to the effective spatial mass
density distribution such as 
\begin{equation}
\rho _{m}^{\ast }\left( x\right) =m_{\Psi }\left\vert \Psi \left( x,t\right)
\right\vert ^{2}  \label{eq14}
\end{equation}%
subject to the normalization condition. Given this, the experienced
gravitational potential is 
\begin{equation}
V^{G}\left( x\right) =-G_{N}\int \frac{\rho _{m}^{\ast }\left( x^{\prime
}\right) }{\left\vert x-x^{\prime }\right\vert }dx^{\prime }  \label{eq15}
\end{equation}%
where $G_{N}$ is the Newton gravitational constant. Thus, the associated
gravitational self-energy reads as 
\begin{equation}
U^{G}=-\frac{1}{2}G_{N}\int \int \frac{\rho _{m}^{\ast }\left( x\right) \rho
_{m}^{\ast }\left( x^{\prime }\right) }{\left\vert x-x^{\prime }\right\vert }%
dxdx^{\prime }  \label{eq16}
\end{equation}%
being the relevant measure of the required breakdown of the quantum
superposition. This gravitational self-energy of the spatially superposed
state $\left\vert \Psi \right\rangle $ could be interpreted as the work
required to bring the $n$ state components $\left\vert \psi
_{i}\right\rangle $ from infinity and superpose them in the required
formation corresponding to the initial state $\left\vert \Psi \right\rangle
. $ In this view, and in terms of point-like mass densities, we can write 
\begin{equation}
\rho _{m}^{\ast }\left( x\right) =\sum\limits_{i=1}^{n}\rho _{i}^{m}\left(
x\right) \text{, \ \ }\rho _{i}^{m}\left( x\right) =m_{\Psi }\delta
^{3}\left( x-x_{i}\right) \text{ }  \label{eq17}
\end{equation}%
from which the gravitational potential energy (\ref{eq16}) could be
expressed in terms of $n$ interacting component state masses as 
\begin{equation}
U^{G}=U_{\Psi }^{G}=-G_{N}\sum\limits_{i,j=1,\ldots ,n}^{j<i}\frac{m_{\psi
_{i}}^{\ast }m_{\psi _{j}}^{\ast }}{\left\vert x_{j}-x_{i}\right\vert }
\label{eq18}
\end{equation}%
where $m_{\psi _{i}}^{\ast }$ and $m_{\psi _{j}}^{\ast }$ are the effective
masses corresponding to the gravitationally-interacting states $\left\vert
\psi _{i}\right\rangle $ and $\left\vert \psi _{j}\right\rangle $ located at
positions $x_{i}$ and $x_{j}$. As it has been mentioned previously \cite%
{20,30}, it has been argued that such a gravitational self-energy arising
from the mismatch of the spacetime geometries, emerged from the mass
energies of the localized states, leads to an instability of the quantum
state (\ref{eq7}) through the suppression of superpositions of the different
spacetime geometry states. More precisely, the exact energy of the breakdown
of the spatial quantum superposition turns out to be the difference between
gravitational self-energies before and after the reduction of the initial
state 
\begin{equation}
\Delta U_{\Psi \rightarrow \psi _{j}}^{G}=U_{\psi _{j}}^{G}-U_{\Psi
}^{G}=-U_{\Psi }^{G}=G_{N}\sum\limits_{i,j=1,\ldots ,n}^{j<i}\frac{m_{\psi
_{i}}^{\ast }m_{\psi _{j}}^{\ast }}{\left\vert x_{j}-x_{i}\right\vert }
\label{eq19}
\end{equation}%
where $U_{\psi _{j}}^{G}=0$ is taken in the sens that the resulting
localized single state $\left\vert \psi _{k}\right\rangle \left\vert
G_{k}\right\rangle $ no more undergoes a gravitational interaction from the
suppressed $n-1$ states. This energy measures, in gravitational terms, the
largeness of the superposition. A likewise physical quantity to be
considered is the electrostatic self-interaction energy that should be also
taken into account in the case of a charged superposed state. Similarly, the
effective spatial charge density distribution $\rho _{Q}\left( x\right) $ is
as follows 
\begin{equation}
\rho _{Q}^{\ast }\left( x\right) =Q_{\Psi }\left\vert \Psi \left( x,t\right)
\right\vert ^{2}  \label{eq20}
\end{equation}%
where $Q_{\Psi }$ is the charge of the state $\left\vert \Psi \right\rangle $%
. Thus, the experienced electrostatic potential reads 
\begin{equation}
V^{E}\left( x\right) =K_{E}\int \frac{\rho _{Q}^{\ast }\left( x^{\prime
}\right) }{\left\vert x-x^{\prime }\right\vert }dx^{\prime }  \label{eq21}
\end{equation}%
where we have used $K_{E}=\left( 4\pi \varepsilon _{0}\right) ^{-1}$ being
the electrostatic constant. In an analogous manner, we can show that the
electrostatic potential energy could be expressed in terms of the $n$
charged interacting localized states as follows 
\begin{equation}
U_{\Psi }^{E}=K_{E}\sum\limits_{i,j=1,\ldots ,n}^{j<i}\frac{Q_{\psi
_{i}}^{\ast }Q_{\psi _{j}}^{\ast }}{\left\vert x_{j}-x_{i}\right\vert }.
\label{eq22}
\end{equation}%
This electrostatic self-interaction energy is expected to be repulsive in
the case of a charged state $\left\vert \Psi =e^{-},p^{+},\ldots
\right\rangle $. Thus, it will have tendency to lower the state decay
process of a quantum state as it resists to its gravitational
self-interaction (\ref{eq16}). This seemingly valuable total self-energy $%
U_{\Psi }^{Self}=U_{\Psi }^{E}+U_{\Psi }^{G}$, which is disregarded in the
standard QMcal dynamics of the wavefunction, should be taken into account
within the Schr\"{o}dinger equation (\ref{eq1}) via an extended potential
such as 
\begin{eqnarray}
i\hbar \triangledown _{t}\Psi \left( x,t\right) &=&\left[ -\frac{\hbar }{%
2m_{\Psi }}\triangledown _{x}^{2}+V_{\Psi }^{Self}+V^{Ext}\right] \Psi
\left( x,t\right)  \notag \\
&=&\left[ -\frac{\hbar }{2m_{\Psi }}\triangledown _{x}^{2}+\left( V_{\Psi
}^{E}+V_{\Psi }^{G}\right) +V^{Ext}\right] \Psi \left( x,t\right)  \notag \\
&=&\left[ -\frac{\hbar }{2m_{\Psi }}\triangledown _{x}^{2}+\left(
K_{E}Q_{\Psi }^{2}-G_{N}m_{\Psi }^{2}\right) \int \frac{\left\vert \Psi
\left( x^{\prime },t\right) \right\vert ^{2}}{\left\vert x-x^{\prime
}\right\vert }dx^{\prime }+V^{Ext}\right] \Psi \left( x,t\right) .
\label{eq23}
\end{eqnarray}%
This could have clearly relevance in the full underlying theory of the state
reduction process that we have not actually use its specified mathematical
details.

\subsection{Collapse time and detectability aspects}

With all the above considerations, and owing to the existing difficulty with
the local gravitational energy concept even in classical GR along with the
argued typical decay time in analogy with the one of an unstable particle,
the time of the state collapse could be approached, in a similar spirit of
the Diosi-Penrose model \cite{19,20,30}, by a representative lifetime of the
general order of 
\begin{equation}
\tau _{\Psi }^{Collapse}\simeq \frac{\hslash }{\left\vert U_{\Psi
}^{G}-U_{\Psi }^{E}\right\vert }  \label{eq24}
\end{equation}%
which is scaled here, not only by the system gravitational potential energy,
but by the difference between the gravitational potential energy and the
eventual electrostatic one of the system. This could slightly impact the
corresponding decay rate since, as mentioned previously, it will have
tendency to lower the decay process of the system as it resists to its
gravitational self-interaction. The collapse time (\ref{eq24}) is, of
course, a very short lifetime. Concretely, by taking the distance $%
\left\vert x-x^{\prime }\right\vert $ as a characterizing superposition
length scale $\ell =:\left\vert x-x^{\prime }\right\vert $ and the effective
mass product of the order $m_{\psi _{i}}^{\ast }m_{\psi _{j}}^{\ast }$ $\sim
m_{\Psi }^{2}$, it can be shown that such a time scale goes as $\sim \ell
/m_{\Psi }^{2}$, which indeed decreases for increasing large mass particles
as the gravitational self-attraction between the localized states increases.
Taking the values of the known gravitational $G_{N}\sim
10^{-11}m^{3}Kg^{-1}s^{-2}$ and electrostatic $K_{E}\sim 10^{10}Nm^{2}C^{-2}$
constants, we can get 
\begin{equation}
\tau _{\Psi }\simeq \hslash d\left\vert \frac{1}{G_{N}m_{\Psi }^{2}}-\frac{1%
}{K_{E}Q_{\Psi }^{2}}\right\vert \sim 10^{-24}\frac{\ell }{m_{\Psi }^{2}}.
\label{eq25}
\end{equation}%
A further ingredient of possible importance is the effective electric
current $I_{\Psi }^{\ast }$ induced by the moving effective charges $Q_{\psi
_{i}}^{\ast }$ during the state reduction process time $\tau _{\Psi }$. For
that, assuming, for simplicity, a steady flow of the moving effective total
charges $\sum\limits_{i=1}^{n}Q_{\psi _{i}}^{\ast }$ transferred through a
certain spatial surface over the collapse time $\tau _{\Psi }$ and the
effective charge product such as $Q_{\psi _{i}}^{\ast }Q_{\psi _{j}}^{\ast }$
$\sim Q_{\Psi }^{2}$, it can be shown that such an induced current goes as $%
\sim m_{\Psi }^{2}/\ell $. Concretely, the corresponding induced effective
electric current can be approached, roughly, as

\begin{equation}
I_{\Psi }^{\ast }\simeq \frac{1}{\tau _{\Psi }}\sum\limits_{i=1}^{n}Q_{\psi
_{i}}^{\ast }\sim 10^{24}Q_{\Psi }\frac{m_{\Psi }^{2}}{\ell }  \label{eq26}
\end{equation}%
which also seems to be of an extremely small intensity. Now, with all these
at hand, and for a certain superposition distance $\ell $, one can now
approximately obtain the decay time and the induced current of the quantum
state reduction process. For simplicity reasons, we restrict to the case of
a superposition of two spatially distinct states $i=1,2$ in a moderate sized
spatial separation distance $\ell \sim 1\mu m$ for different mass/complexity
scales (number of constituent: electrons, protons and neutrons)
corresponding to some physical objects. For that, we estimate, for
comparison, the collapse time and the induced current of the state reduction
for certain systems where we have used for the case of massless states,
i.e., the case of photon, the corresponding kinetic mass $m_{\gamma }\equiv
hc^{-1}\lambda _{visible}^{-1}$ of the visible light range $\sim 10^{-11}m$.
In table 1, we collect the following results:

\begin{center}
\bigskip 
\begin{tabular}{|l|l|l|l|l|l|l|l|}
\hline
State: $\left\vert \Psi \right\rangle $ & $\left\vert \gamma \right\rangle $
& $\left\vert e\right\rangle $ & $\left\vert H\right\rangle $ & $\left\vert
H_{2}O\right\rangle $ & $\left\vert ...\right\rangle $ & $\left\vert
Ant\right\rangle $ & $\left\vert Cat\right\rangle $ \\ \hline
Number of particles: $N_{p}$ & $1$ & $1$ & $2$ & $28$ & $...$ & $\sim
10^{23} $ & $\sim 10^{29}$ \\ \hline
Mass: $m_{\Psi }\left( kg\right) $ & $\sim 10^{-49}$ & $10^{-32}$ & $%
10^{-27} $ & $10^{-26}$ & $...$ & $10^{-6}$ & $4$ \\ \hline
Charge: $Q_{\Psi }\left( C\right) $ & $0$ & $-1$ & $0$ & $0$ & $...$ & $0$ & 
$0$ \\ \hline
Collapse time: $\tau _{\Psi }\left( s\right) $ & $10^{70}$ & $10^{35}$ & $%
10^{25}$ & $10^{23}$ & $...$ & $10^{-17}$ & $10^{-30}$ \\ \hline
Electric current: $I_{\Psi }^{\ast }\left( A\right) $ & $0$ & $10^{-35}$ & $%
0 $ & $0$ & $...$ & $0$ & $0$ \\ \hline
\end{tabular}

Table 1: Collapse time of the quantum state of certain physical systems.
\end{center}

From this table, we can observe the huge hierarchy in the collapse time
between the microscopic and the macroscopic states. 
% $\tau _{\Psi _{molecule}}\sim
%10^{40}\tau _{\Psi _{in\sec t}}$. 
This means that when atoms glue together to form larger systems, the
superposition principle, being the cornerstone of quantum theory,
dramatically breaks down. A possible test of the collapse time requires
making large superposition $\ell \gg 1\mu m$ of a massive system to ensure
the collapse time is measurable enough for the state reduction process
before any type of external noise perturbs the measurement. However, the
main obstacle in performing this and analogous suggestions is to keep it
stable for times comparable to $\tau _{\Psi }$. This is what justifies the
difficulty of observing quantum superposition at classical scales, unless
ultra time-sensitive devices that can be used to capture and study such
ultra rapid processes are available. Hopefully, since the ability to observe
events on such extremely tiny timescales is important for basic physics,
i.e., to understand how particles behave and how atoms move within
molecules, physicists are continuously pushing the time limits ever further
towards going over the recently attained attosecond limit $\sim 10^{-18}s$ 
\cite{33}, ultimately zeptosecond $\sim 10^{-21}s$ see even yoctosecond
ranges $\sim 10^{-24}s$. Achieving that necessitates driving technology to
realize pulses using higher-wavelength sources, and also producing pulses
that encompass a wider range of frequencies. Regarding the induced current,
the same challenge is faced. Measuring such an extremely small intensity
requires ultra-sensitive multimeters with measurement capability beyond the
smallest ranges recently reached.

With regard to present-day experimental data and the possible constraints
that can be put on the proposed model, we believe that this could be done in
an indirect way by exploiting an unavoidable side effect of the induced
electric current that the collapse of a chagred state could come with. More
precisely, this could be supported by means of the possible emission of the
associated EM radiation with a spectrum that depends on the configuration of
the system (\ref{eq26}). In such a case, taking the wavelength of this
produced EM radiation to be of the order of the wavelength state $\lambda
_{\gamma }\sim \lambda _{\Psi }=\frac{h}{m_{\Psi }v}$, where $v$ is the
system velocity, i.e. $v=\frac{\mathbf{\ell }}{\tau _{\Psi }}$, \ the
corresponding energy of the associated radiation is $E_{\gamma }\sim \frac{%
m_{\Psi }\ell c}{\tau _{\Psi }}.$ Although such a predicted radiation energy
appears to be faint for potential detection even for the existing
experiments performed in very low-noise environments \cite{34,35}, we think
that particular choices of the spatial separation distance $\ell $ could
lead, at least, to a constraint setting on some involved parameters of the
proposed model. 

\section{Discussion and further considerations}

In this work, we have considered the possible implication of gravity in the
reduction of the quantum state within a physical wavefunction approach. More
precisely, it has been shown that the gravitational self-interaction energy
emerged from the superposed spacetime curvature states created by the
effective masses of the localized eigenstates leads to the decay of the
quantum state into one of its localized states, during a certain interaction
time. We have described such a gravity-driven state reduction process and
established the emerged gravitational self-energy of the superposed state,
and approached the corresponding lifetime. We have pointed out that the
obtained extremely small lifetime for macroscopic states seems to lie beyond
the reach of the present as well as the near future timescales observation
abilities. For charged states, we have approached the effective electric
current which could be induced during the state reduction. It has been also
appeared to be of an extremely small intensity necessitating
highly-sensitive electric current measuring devices with measurement
capability going far beyond the smallest ranges recently reached. Thus,
despite the fact that these results seem reasonable, but for the moment they
appear to be rather beyond what can be experimentally tested. Unfortunately,
the approach described in this proposal does not provide a full theory of
quantum state reduction necessitating deeper mathematical considerations,
but it solely points out the level at which deviations from standard quantum
state concept are likely to be considered owing to gravitational effects.
Such mathematical considerations would have to originate from quite other
related directions, mainly quantum field theory and quantum gravity. A
recent example in this direction is the recently proposed alternative
approach towards the construction of a consistent theory of classical
gravity coupled to quantum field theory \cite{36}, where the interaction of
the classical spacetime with the quantum degrees of freedom inevitably leads
to decoherence in the quantum system. Therefore, one gets by the classical
metric a fundamental decoherence of the quantum field. Such an alternative
classical quantum theory can be viewed as fundamental or as an effective
theory advantageous for dealing with the back-reaction of quantum fields on
geometry. Thus, in the spirit of such quantum classical interaction theories
including our proposed approach, the classicality of the system is not an
emergent or effective property of largeness, but rather, fundamental, as it
is inherited from spacetime. A consequence of this is that the QMcal
measurement postulate is no longer required. Thus, \ the expectation that no
completely satisfactory theory will be upcoming until there is a scientific
revolution in the description of quantum phenomena which remains more likely.%
\textbf{\ }

\textbf{\bigskip }


\begin{thebibliography}{99}
\bibitem{1} M. Born, "On the Quantum Mechanics of Collision Processes".
Zeitschrift f\"{u}r Physik. 37 (12): 863--867 (1926).

\bibitem{2} R. Feynman, R. Leighton, M. Sands, "The Feynman Lectures on
Physics". Vol. 3. California Institute of Technology. ISBN 978-0201500646
(1964).

\bibitem{3} G. Jaeger, "What in the (quantum) world is macroscopic?",
American Journal of Physics. 82 (9): 896--905 (2014).

\bibitem{4} A. Einstein, B. Podolsky, N. Rosen, "Can quantum-mechanical
description of physical reality be considered complete?", Phys. Rev. 47,
777--780 (1935).

\bibitem{5} D. Bohm, "Quantum Theory", Prentice-Hall, Upper Saddle River
(1951).

\bibitem{6} J. S. Bell, "On the problem of hidden variables in quantum
theory", Rev. Mod. Phys. 38, 447--452 (1966).

\bibitem{7} E. Schrodinger, "Die gegenwartige Situation in der
Quantenmechanik", Naturwissenchaften 23, 807-812, 823-828, 844-849 (1935).

\bibitem{8} J. von Neumann, "Mathematical Foundations of Quantum Mechanics",
Princeton Uni versity Press (1955).

\bibitem{9} J. D. Trimmer, "The present situation in quantum mechanics: a
translation of Schr%
%TCIMACRO{\U{a8}}%
%BeginExpansion
\"{}%
%EndExpansion
odinger's cat paradox paper", Proc. Amer. Phis. Soc. 124, 323-338 (1980).

\bibitem{10} J. A. Wheeler and W.H. Zurek, "Quantum theory of measurement",
Princeton series in physics (1983).

\bibitem{11} M. Jammer, "The conceptual development of quantum mechanics",
MCGraw-Hill (1966).

\bibitem{12} M. F. Pusey, J. Barrett, T. Rudolph, "On the reality of the
quantum state", Nature Physics. 8 (6): 475--478 (2012).

\bibitem{13} H. Everett et al., "The Many-Worlds Interpretation of Quantum
Mechanics", Princeton Series in Physics. Princeton, NJ: Princeton University
Press. p. v. ISBN 0-691-08131-X (1973).

\bibitem{14} G. Ghirardi, A. Rimini and T. Weber, "Unified dynamics for
microscopic and macroscopic systems", Phys. Rev. D 34, 470-491 (1986).

\bibitem{15} P. Pearle, "Combining stochastic dynamical state-vector
reduction with spontaneous localization", Phys. Rev. A 39, 2277-2289 (1989).

\bibitem{16} G. Ghirardi, P. Pearle and A. Rimini, "Markov processes in
Hilbert space and continuous spontaneous localization of systems of
identical particles", Phys. Rev. A 42, 78-89 (1990).

\bibitem{17} A. Bassi and G. Ghirardi, "Dynamical reduction models", Phys.
Rep. 379, 257-426 (2003).

\bibitem{18} A. Bassi, K. Lochan, S. Satin, T. Singh and H. Ulbricht,
"Models of wave function collapse, under lying theories, and experimental
tests", Rev. Mod. Phys. 85, 471-527 (2013).

\bibitem{19} L. Di\'{o}si, "Models for universal reduction of macroscopic
quantum fluctuations". Physical Review A. 40 (3): 1165--1174 (1989).

\bibitem{20} R. Penrose, "On Gravity's role in Quantum State Reduction".
General Relativity and Gravitation. 28 (5): 581--600 (1996).

\bibitem{21} J. S. Bell, "Against 'measurement'", Physics World 3, 33 (1990).

\bibitem{22} H. Everett, ""Relative State" Formulation of Quantum
Mechanics", Phys.Rev.Mod., 29, 454, (1957).

\bibitem{23} L. de Broglie, "Tentative d'Interpretation Causale et
Nonlineaire de la Mechanique Ondulatoire" (Gauthier-Villars, Paris) (1956).

\bibitem{24} D. Bohm, "A Suggested Interpretation of the Quantum Theory in
Terms of "Hidden" Variables. I", Phys. Rev. 85, (1952).

\bibitem{25} D. Bohm, and B. Hiley, "The Undivided Universe" (Routledge,
London) (1994).

\bibitem{26} M. Gell-Mann, and J.B. Hartle, "Classical equations for quantum
systems, Phys. Rev. D47, 3345 (1993).

\bibitem{27} R. Griffiths, "Consistent Histories and the Interpretation of
Quantum Mechanics", J. Slat. Phys. 36, 219 (1984).

\bibitem{28} R. Hang, "Local Quantum Physics: Fields, Particles, Algebras"
(Springer- Verlag, Berling). (1992).

\bibitem{29} R. Omnes, "Consistent interpretations of quantum mechanics",
Rev. Mod. Phys. 64, 3 (1992).

\bibitem{30} R. Penrose, "On the Gravitization of Quantum Mechanics 1:
Quantum State Reduction". Found Phys 44:557-575, (2014).

\bibitem{31} A.J. Leggett, Macroscopic quantum systems and the quantum
theory of measurement. Prog. Theor.Phys.Suppl. 69,80--100(1980).

\bibitem{32} S. Weinberg, "Collapse of the state vector", Phys.Rev.A
85,062116(2012).

\bibitem{33} F. Krausz and M. Ivanov, "Attosecond physics", Rev. Mod. Phys.
81, 163 (2009).

\bibitem{34} S. Donadi, K. Piscicchia, C. Curceanu, L. Di\'{o}si, M.
Laubenstein, A. Bassi, "Underground test of gravity-related wave function
collapse", Nature Physics 17, 74 (2021).

\bibitem{35} N.-H. Kaneko, T. Tanaka1 and Y. Okazaki, "Perspectives of the
generation and measurement of small electric currents", Meas. Sci. Technol.
35 011001 (2024).

\bibitem{36} J. Oppenheim, "A Postquantum Theory of Classical Gravity"
Physical Review X 13 (4), 041040 (2023)."
\end{thebibliography}
\end{document}